\documentclass{aa}

\usepackage{graphicx}
\usepackage{xcolor}
\usepackage{txfonts}
\usepackage{gensymb}
\usepackage{textgreek}
\usepackage{color}
\usepackage[colorlinks=true]{hyperref}
\hypersetup{linkcolor=blue,citecolor=blue,filecolor=blue,urlcolor=blue}
\begin{document} 

   \title{Polarimetric Diversity in Tidal Disruption Events: Comparative Study of Low-Polarised sources with AT2020mot}

\author{A. Floris \inst{1,2,3}
  \and I. Liodakis \inst{1,4}
  \and K. I. I. Koljonen \inst{5}
  \and E. Lindfors \inst{6,7}
  \and B. Ag\'is-Gonzalez\inst{1}
  \and A. Paggi \inst{1,2}
  \and D. Blinov \inst{1,2}
  \and K. Nilsson \inst{7}
  \and I. Agudo \inst{8}
  \and P. Charalampopoulos \inst{6}
  \and M. A. D\'iaz Teodori \inst{6,9}
  \and J. Escudero Pedrosa \inst{8,10}
  \and J. Otero-Santos \inst{8,11}
  \and V. Piirola \inst{6}
  \and M. Newsome \inst{12,13}
  \and S. Van Velzen \inst{14}
}

    \institute{Institute of Astrophysics, FORTH, N.Plastira 100, Vassilika Vouton, 70013 Heraklion, Greece\\  \email{afloris@ia.forth.gr}
    \and Department of Physics University of Crete, Voutes University Campus, 70013 Heraklion, Greece
    \and National Institute for Astrophysics (INAF), Astronomical Observatory of Padova, IT-35122 Padova, Italy
    \and NASA Marshall Space Flight Center, Huntsville, AL 35812, USA
    \and Institutt for Fysikk, Norwegian University of Science and Technology, Høgskloreringen 5, Trondheim, 7491, Norway
    \and Department of Physics and Astronomy, 20014 University of Turku, Finland
    \and Finnish Centre for Astronomy with ESO (FINCA), Quantum, Vesilinnantie 5, 20014 University of Turku, Finland
    \and Instituto de Astrofísica de Andalucía, IAA-CSIC, Glorieta de la Astronomía s/n, 18008 Granada, Spain
    \and Nordic Optical Telescope, Rambla José Ana Fernández, Pérez 7, E-38711 Breña Baja, Spain
    \and Center for Astrophysics \textbar ~Harvard \& Smithsonian, 60 Garden Street, Cambridge, MA 02138 USA
    \and Istituto Nazionale di Fisica Nucleare, Sezione di Padova, 35131 Padova, Italy
    \and Las Cumbres Observatory, 6740 Cortona Drive, Suite 102, Goleta, CA 93117-5575, USA
    \and Department of Astronomy, The University of Texas at Austin, 2515 Speedway, Stop C1400, Austin, TX 78712, USA
    \and Leiden Observatory, Leiden University, PO Box 9513, 2300 RA Leiden, The Netherlands
    }

   \date{}

 
  \abstract
   {Tidal disruption events (TDEs) occur when a star is disrupted by the tidal forces of a supermassive black hole (SMBH), producing bright multi-wavelength flares. Among these events, AT2020mot has so far exhibited the highest recorded optical polarisation, with tidal shocks proposed as the primary source of its polarised emission.}
   {We present a comprehensive analysis of 13 TDEs with available polarimetric observations, aiming to determine whether the unusually high polarisation of AT2020mot stems from unique physical processes or arises from mechanisms shared by other TDEs.}
   {We present new optical polarisation measurements of TDEs obtained from multiple ground-based telescopes, combining them with optical, UV, and X-ray light curves from the Zwicky Transient Facility and the \textit{Swift} observatory, respectively. We derive intrinsic TDE properties—such as SMBH and stellar masses—using {\tt MOSFiT} and {\tt TDEMass}, and compare them with the ones of the sample population.}
   {Our population study reveals that AT2020mot agrees with the broader TDE sample in most physical properties, including blackbody temperature, luminosity, and rise timescales. However, its optical polarisation degree is exceptionally high compared to the low or undetected polarisation observed in other events. Additionally, AT2020mot appears to have an elevated column density from our {\tt MOSFiT} fits, suggesting a more complex environment than is typically assumed.}
   {We conclude that although AT2020mot fits well within the general TDE population in terms of global characteristics, its extraordinarily high polarisation and higher column density challenge current models based purely on shock or reprocessing mechanisms. More extensive, time-resolved polarimetric monitoring of newly discovered TDEs will be critical to determine whether AT2020mot represents an outlier or the extreme end of a continuum of TDE properties.}


   \authorrunning{A. Floris et al.}
   \titlerunning{Polarimetric diversity in Tidal Disruption Events}
   \maketitle

\section{Introduction}
\label{intro}

Tidal disruption events (TDEs) are transient astrophysical phenomena that occur when a star is scattered onto an orbit passing sufficiently close to the supermassive black hole (SMBH) in the central region of its host galaxy \citep{rees1988, komossa2015}. Once the black hole's tidal forces overcome the star's self-gravity, stripping the star of its gas \citep{hills1975}, a bright flare is produced that typically emits across the X-ray, ultraviolet (UV), optical and infrared (IR) wavelengths.

UV–optical TDEs are identified through pronounced flares in these bands and by their broad optical and UV spectral lines \citep{vanvelzen2020}. In most cases, however, the optical peak is not accompanied by contemporaneous X‑ray emission \citep{gezari2008,gezari2012,holoien2016a,auchettl2018,hinkle2021,vanvelzen2021}. The origin of the optical luminosity, and the apparent absence of contemporaneous X-rays, therefore remains debated. Two main scenarios have been proposed to explain these observations: (1) an accretion disk forms rapidly after disruption and produces X-rays, which are then reprocessed into UV–optical emission by an optically thick layer of gas at radii much larger than the tidal radius \citep{metzger2016,dai2018}, and (2) a predominantly shock-powered mechanism, where stellar debris streams collide in the outer regions of a highly eccentric disk, rather than power generating primarily from accretion onto the black hole \citep{piran2015,shiokawa2015}. Different events appear to favour different physical mechanisms. For example, accretion disk models are often invoked to explain Bowen fluorescence lines \citep{leloudas2019,blagorodnova2019} and coronal emission lines \citep{trakhtenbrot2019,koljonen2024}, presumably formed by excitation from high-energy photons, even in TDEs where X-rays are partially or completely obscured. On the other hand, the variable polarisation reported in several TDEs \citep{leloudas2022,patra2022,liodakis2023,koljonen2024} and, in particular, the extraordinarily high polarisation degree observed in AT2020mot (the highest measured to date in the absence of a jet, $\Pi \sim 25 \pm 4\%$; \citealt{liodakis2023}), strongly supports that at least some TDEs involve mechanisms different from electron-scattering. Pure electron-scattering reprocessing scenarios generally predict a maximum polarisation degree $<14\%$ \citep{leloudas2022,charalampopoulos2023}.

Polarimetry offers a sensitive probe of TDE geometry and emission physics. Even in the absence of intrinsically polarised sources (e.g. relativistic jets, which appear to be rare among optical TDEs; \citealt{wiersema2012,wiersema2020}), moderate to high polarisation can arise through electron scattering whenever the reprocessing medium or shock fronts are aspherical. Measurements of the polarisation degree ($\Pi$) and polarisation angle ($\Theta$) thus constrain the shape and orientation of the scattering region \citep{chornock2014,patra2022}, and can discriminate between reprocessing-dominated and shock-powered scenarios.

In this context, AT2020mot stands out as a unique source: all other TDEs with polarimetric measurements exhibit significantly lower $\Pi$, even before accounting for its host-galaxy contamination \citep{liodakis2023}. 
Since the integrated host galaxy emission is unpolarized it effectively depolarizes the TDE emission measured within a circular aperture. Therefore, the host galaxy correction renders higher level of polarization. Late-time polarimetry, obtained after the flare has faded, is therefore indispensable for isolating the intrinsic polarisation of the event \citep{andruchow2008}.

Here we present new optical polarimetric observations obtained within the Black hOle Optical polarization TimE-domain Survey (BOOTES) programme, a campaign designed to systematically monitor TDE polarisation. We report our first results, including late-time measurements of AT2020mot, and perform a homogeneous analysis that combines these data with photometry for a sample of well-observed TDEs. Our goal is to compare the multiwavelength properties of AT2020mot to the ones of other TDEs in order to determine the physical origin for why it stands out among the rest of the TDE population.

We describe our sample selection and observations in Section \ref{obs}, including new polarisation measurements for the sources in our sample. Section \ref{res} outlines the methods used to estimate TDE properties and the results of our population study, and we discuss the interpretation of our findings in Section \ref{discussions}. Finally, we provide a summary of our conclusions in Section \ref{conclusions}.

Throughout this paper, we adopt a flat $\Lambda$CDM cosmology with $H_0 = 67.4\ \mathrm{km\ s^{-1}\ Mpc^{-1}}$, $\Omega_\mathrm{M}=0.315$, and $\Omega_\Lambda=0.685$  \citep{planck2020}.

\section{New polarisation observations}
\label{obs}

\subsection{Sample}

The sample used in this work consists of TDEs that were monitored as part of the BOOTES program of polarimetric observation, prior to mid-2024. The data were gathered as part of an ongoing campaign to track the polarisation evolution of TDEs with several ground-based facilities, each source being observed at multiple epochs throughout its flare and subsequent decline.

The polarimetric observations were obtained with the following instruments:
\begin{itemize} 
\item The RoboPol polarimeter \citep{ramaprakash2019} on the 1.3 m telescope at the Skinakas Observatory, Crete; 
\item The ALFOSC polarimeter \citep[see][for details on data reduction and calibration]{nilsson2018} on the Nordic Optical Telescope (NOT); 
\item The DIPOL-1 polarimeter \citep{oterosantos2024} on the 90 cm telescope (T90) at the Observatorio de Sierra Nevada (OSN);
\item The CAFOS polarimeter\footnote{\url{https://www.caha.es/telescope-2-2m/cafos}} \citep[see][for details on data reduction and calibration using CAFOS and DIPOL-1]{escuderopedrosa2024} on the 2.2 m telescope at the Centro Astronómico Hispano en Andalucía (CAHA).
\end{itemize}

Standard polarimetric calibration procedures were applied to all data. We observed both unpolarised and polarised standard stars to verify the accuracy of the instrumental polarisation and to calibrate the instrumental zero-point for the polarisation angle. Uncertainties in the polarisation measurements are derived through error propagation including photon noise of the detector, background subtraction and instrumental corrections for both $\Pi$ and $\Theta$ from standard stars.

Typical exposure times per target ranged from a few minutes up to over an hour, depending on the source magnitude and the specific telescope/instrument configuration in order to obtain a sufficiently high signal-to-noise ratio (S/N) that ensured reliable polarisation measurements or 3$\sigma$ upper limits. 

The new measurements reported here are listed in Table~\ref{tab:newpol}; previously published polarimetry for AT2020mot \citep{liodakis2023}, AT2022fpx \citep{koljonen2024}, and AT2023clx \citep{koljonen2025} are also included.

The final sample comprises 13 TDEs. Among these, AT2020mot is the only source displaying a high polarisation degree (\mbox{$\Pi \approx 25\%$}), posing a challenge to current models. By contrast, the remaining TDEs exhibit polarisation degrees of $\Pi < 6\%$ or yield only non-detections ($\Pi - 3\sigma_{\Pi} < 0\%$, where $\sigma_{\Pi}$ is the associated uncertainty). One of the TDEs in our sample, AT2022dbl, has undergone two flares: the first in February 2022 \citep{arcavi2022dbl} and the second in late 2023 \citep{yao2024}, which have been interpreted as arising from partial disruption of the same star \citep[pTDE;][]{lin2024}. In this work we will treat the two flares separately, indicating the second flare as "AT2022dbl$_2$".

Basic properties of the full sample, together with the maximum observed polarisation degree ($\Pi_{\mathrm{max}}$), are summarised in Table~\ref{tab:sample}.

\begin{table}[h!]
\renewcommand{\arraystretch}{1.25}
\caption{TDE sample properties}
\label{tab:sample}
\centering
\begin{tabular}{l c c c}
\hline\hline
Name & Spectral type & z & $\Pi_{\rm max}$\\
 & & & [\%]\\
(1) & (2) & (3) & (4) \\
\hline
AT2020mot & TDE H+He & 0.07 & $25\pm4$\\ 
AT2020afhd & TDE H+He & 0.027 & $<2.8$\\ 
AT2022dbl & TDE H+He & 0.0284 & $<2.1$\\ 
AT2022fpx & TDE H+He + ECLE & 0.073 & $2.4\pm0.5$\\ 
AT2022gri & TDE featureless & 0.028 & $1.29\pm0.39$\\ 
AT2022hvp & TDE He & 0.12 & $6.86\pm1.14$\\ 
AT2022upj & TDE He+ECLE & 0.054 & $6.40\pm1.20$\\ 
AT2022wtn & TDE H+He & 0.049 & $<0.72$\\ 
AT2023clx & TDE H+He & 0.01107 & $4.8\pm0.6$\\ 
AT2023ugy & TDE H & 0.106 & $<1.2$\\
AT2023lli & TDE H+He & 0.036 & $0.62\pm0.20$\\ 
AT2024bgz & TDE H+He & 0.0585 & $<8.1$\\ 
AT2024gre & TDE featureless & 0.12 & $4.1\pm1.1$\\ 
\hline
\end{tabular}
\tablefoot{(1) TNS name of the source. (2) Spectral type of the TDE. (3) Redshift from the TNS. (4) Maximum host-corrected polarisation degree of the source detected. Upper limits are 3-$\sigma$ non-detections.}
\end{table}

\subsection{AT2020mot follow-up observations}

As discussed in Section \ref{intro}, it is critical to measure the TDE’s intrinsic polarisation, which requires subtracting any host contribution. In \citet{liodakis2023}, the polarisation degree of AT2020mot was corrected under the assumption that its host galaxy was unpolarised. To verify this assumption, we obtained late-time polarisation observations of the host galaxy, once the flare had faded. We confirm through multiple observations that the host galaxy is unpolarised ($\Pi<0.75\%$ at 3$\sigma$ confidence interval). This confirms that AT2020mot, which was the most polarised TDE to date even in absence of host-correction ($\Pi\sim8.3\%$ uncorrected), had a maximum intrinsic polarisation degree $\sim 25\%$ \citep{liodakis2023}. Figure~\ref{fig:motcurve} illustrates the polarisation evolution of AT2020mot alongside its optical and UV light curves.

\subsection{Host-galaxy correction}
\label{hostcorr}

For sources with at least one significant polarisation detection and a non‑negligible host contribution to the photometry, we corrected for host‑galaxy dilution following \cite{liodakis2023}, assuming that the host-galaxy is unpolarised: 
\begin{equation}
\Pi_{\mathrm{corr}} = \Pi_{\mathrm{obs}}\times\frac{I}{I-I_{\mathrm{host}}}.
\end{equation}
In the previous equation, $\Pi_{\rm obs}$ is the observed polarisation degree, $I$ is the observed total flux density of the source at the time of the polarimetric observation, and $I_{\rm host}$ the host-galaxy flux measured as the median flux obtained from pre-flare observations in the $g$ and $r$ bands from the Zwicky Transient Facility (ZTF). Photometry in the $g$ and $r$ bands was converted to photometry in the filter used for polarimetric observations using the transformations described in \cite{tonry2012}. AT2023lli, despite exhibiting detections of $\Pi$, lacks contemporaneous ZTF photometry, preventing a reliable host correction. All host‑corrected measurements are listed in Table\ref{tab:hostcorr}. 
Additionally, we estimate the maximum galactic interstellar polarisation in the various adopted bands using the \cite{serkowski1975} law and the \cite{schlafly2011} reddening maps. All sight-lines yield a maximum expected interstellar polarisation ($\Pi_{\rm max}$) smaller than a fraction of a percent, well below our statistical errors, thus no interstellar polarisation correction was applied, since it is also negligible with respect to the host-correction.

\subsection{Spectral classification}

Spectral observations are essential both for distinguishing TDEs from other types of transients and for tracing their physical properties over time \citep{vanvelzen2020, holoien2020, charalampopoulos2024}.  We report the spectral type of the sources as observed in their classification spectra available on the Transient Name Server (TNS) website\footnote{\url{https://www.wis-tns.org/}}, and classified according to the convention detailed in \cite{vanvelzen2020,langis2025}:

\begin{itemize}
    \item {\bf H TDEs} exhibit only broad Balmer lines in their optical spectrum;
    \item {\bf He TDEs} exhibit only broad He{\sc{ii}}$\lambda$4686 lines in their optical spectrum;
    \item {\bf H+He TDEs} exhibit both Balmer and He{\sc{ii}} broad lines in their optical spectrum;
    \item {\bf Featureless TDEs} exhibit no broad lines in their optical spectrum.
\end{itemize}

When high-ionisation coronal lines are present we denote the source as an extreme coronal-line emitter (ECLE; \citealt{trakhtenbrot2019}). Approximately half of our sample are H+He TDEs; the remainder consists of one pure H TDE (AT2023ugy), two He TDEs, and two featureless events. Two objects in the sample are classified as ECLEs.

\section{Population study}
\label{res}

Our goal is to identify what makes AT2020mot uniquely highly polarised, whereas the majority of optical TDEs exhibit little or no polarisation.  To this end we compared the UV, optical, and X–ray light curves of the 13 events in our sample and searched for correlations between their photometric and polarimetric properties.

We used photometry from the \textit{Swift} Ultraviolet/Optical Telescope (UVOT; \citealt{Roming2005}), which provides three near-UV bands (UVW2, UVM2, UVW1) and three optical bands (U, B, V). Optical light curves in the ZTF $g$ and $r$ bands were obtained from the BHTOM\footnote{\url{https://bh-tom2.astrolabs.pl/}} archive and processed as described by \cite{langis2025}. X–ray data were collected from all available \textit{Swift-XRT} and \textit{XMM–Newton} observations; the reduction followed the procedure of \cite{langis2025}.  Source‐by‐source X–ray properties are summarised in Appendix~\ref{sampdesc}.

\begin{figure*}
\centering
\includegraphics[width=\hsize]{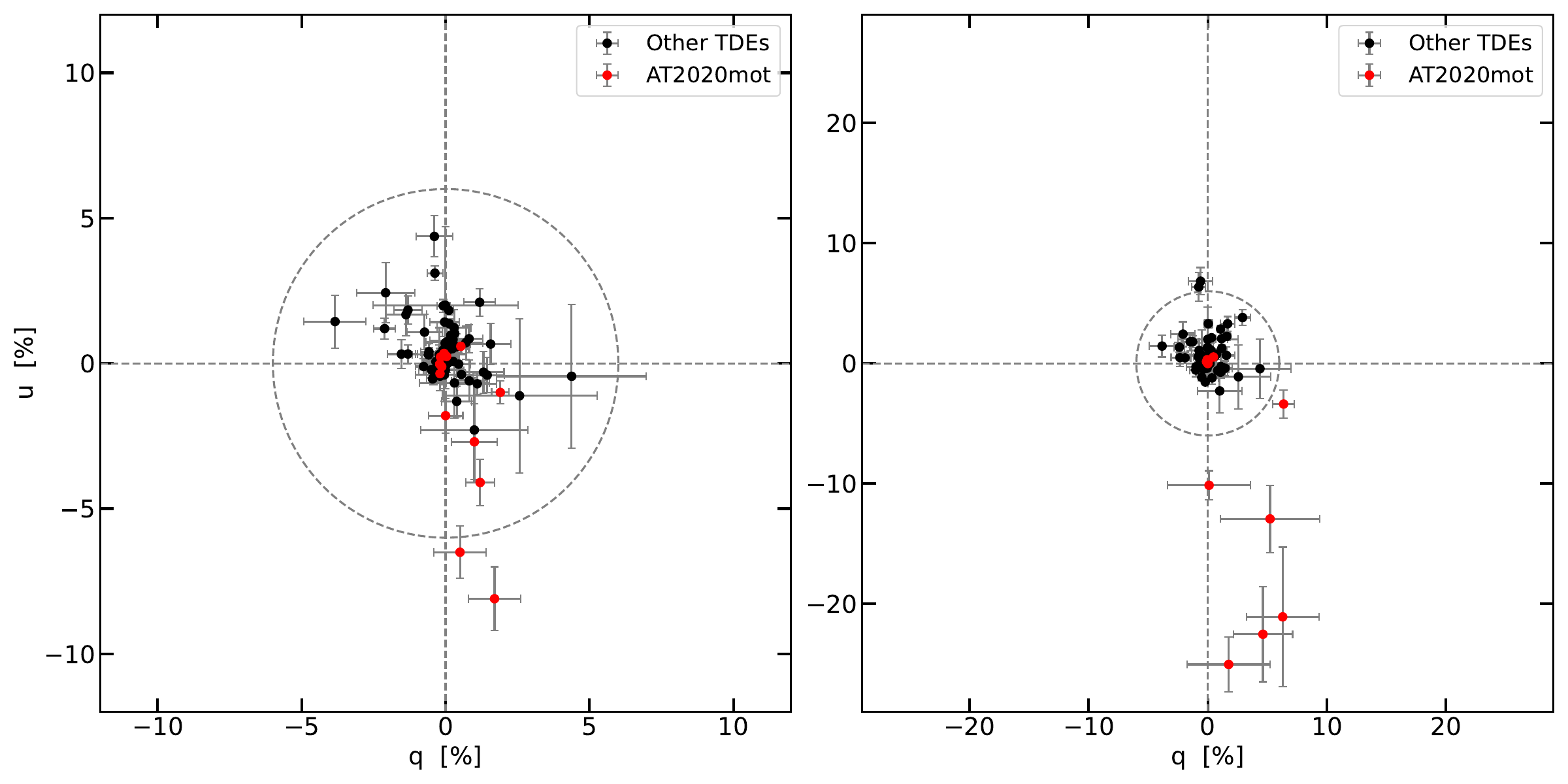}
\caption{{\it Left panel}: Observed Stokes q and u parameters for the sample used in this work. Measurements associated with AT2020mot are highlighted in red. The dashed grey circle represents a value of $\Pi=6\%$. {\it Right panel}: Same, but for the host-corrected Stokes q and u parameters for the sample used in this work.}
\label{fig:quhist}
\end{figure*}

Figure~\ref{fig:motcurve} displays the multi-band light curve of AT2020mot alongside its measured $\Pi$ and $\Theta$. UVOT magnitudes were converted from the Vega system to the AB system using the conversion factors from \cite{blanton2007}. Figure~\ref{fig:quhist} displays the distributions of the Stokes parameters $q$ and $u$, highlighting that AT2020mot is the most highly polarised TDE in our sample. By contrast, the other TDEs are mostly consistent with low polarisation, even after the host-correction. Both near the peak (approximately 23 days after the peak and then again approximately 76 days after peak luminosity) and at late times, AT2020mot also appears unpolarised, matching the general TDE population. This raises some fundamental questions, suggesting that the other TDEs might have been missed at epochs of potentially higher $\Pi$. This prompted us to perform a broader population study, exploring the physical parameters derived from our multi-wavelength dataset.

\begin{figure*}
\centering
\includegraphics[width=\hsize]{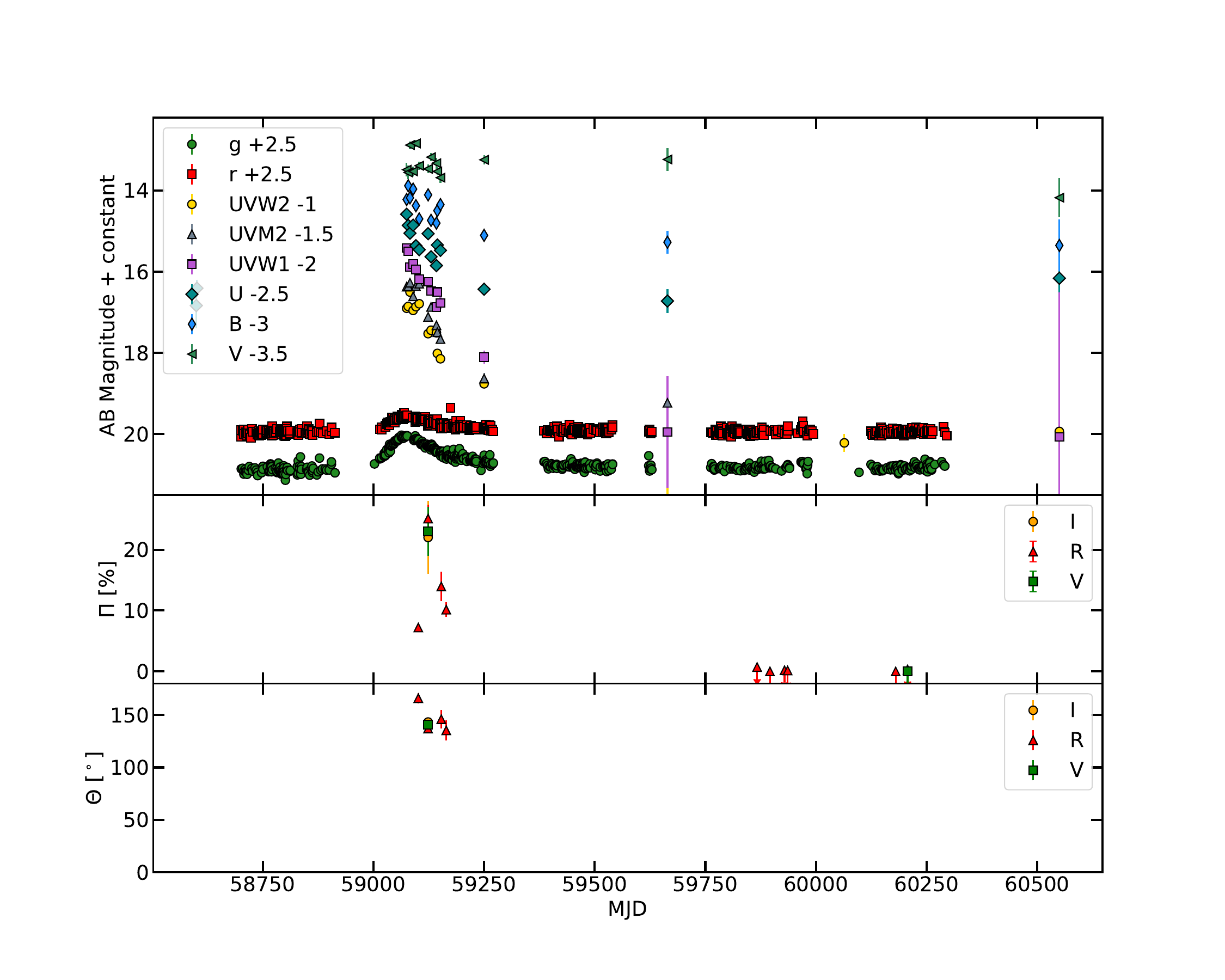}
\caption{{\it Top panel}: AB-magnitude light curve of AT2020mot, rescaled for clarity. {\it Middle panel}: Intrinsic $\Pi$ measurements of AT2020mot over time. Non-detections, defined by $\Pi - 3\sigma_\Pi < 0\%$, are plotted as upper limits. {\it Bottom panel}: $\Theta$ measurements of AT2020mot over time. Non-detections are omitted, as $\Theta$ is undefined in such cases.}
\label{fig:motcurve}
\end{figure*}


\subsection{Blackbody SED fitting}
\label{bbfit}

We modelled the spectral energy distribution (SED) of each TDE at peak brightness with a single black body, using an \texttt{emcee}–based algorithm \citep{emcee2013}. The fits combined host‑subtracted \textit{Swift}/UVOT and ZTF photometry: for UVOT we adopted the latest available post‑flare observations as a host template, while for ZTF we subtracted the median pre‑flare magnitude. All fluxes were shifted to the rest frame and corrected for Galactic extinction using the \citet{schlafly2011} reddening maps.

From the best-fit blackbody temperature and distance $d$ of each TDE, we calculated the peak bolometric luminosity ($L_{\rm bol}$), adopting the cosmology defined in Section \ref{intro}. Table \ref{tab:bolbb} lists the resulting bolometric luminosities and blackbody temperatures.

\begin{table}[h!]
\renewcommand{\arraystretch}{1.25}
\caption{Peak bolometric luminosities and blackbody temperatures}
\label{tab:bolbb}
\centering
\begin{tabular}{l c c}
\hline\hline
Name & $\log L_{\rm bol}$ & $T_{\rm bb}$ \\
 & [erg s$^{-1}$] & [K] \\
(1) & (2) & (3)\\
\hline
AT2020mot & $44.10^{+0.03}_{-0.03}$ & $31706^{+2861}_{-2319}$ \\
AT2020afhd & $44.03^{+0.01}_{-0.02}$ & $47698^{+1641}_{-2599}$ \\
AT2022dbl & $43.81^{+0.02}_{-0.02}$ & $35413^{+2055}_{-1812}$ \\
AT2022dbl$_2$ & $43.36^{+0.03}_{-0.04}$ & $36158^{+4013}_{-3194}$ \\
AT2022fpx & $43.77^{+0.01}_{-0.01}$ & $15187^{+341}_{-323}$ \\
AT2022gri & $43.42^{+0.02}_{-0.02}$ & $22209^{+1331}_{-1135}$ \\
AT2022hvp & $45.29^{+0.01}_{-0.01}$ & $39734^{+1051}_{-1013}$ \\
AT2022upj & $43.73^{+0.03}_{-0.03}$ & $29632^{+2610}_{-2114}$ \\
AT2022wtn & $43.35^{+0.01}_{-0.01}$ & $14912^{+503}_{-459}$ \\
AT2023clx & $42.66^{+0.01}_{-0.01}$ & $15280^{+427}_{-400}$ \\
AT2023ugy & $44.54^{+0.02}_{-0.02}$ & $47987^{+1473}_{-2800}$ \\
AT2023lli & $43.67^{+0.04}_{-0.04}$ & $37242^{+6097}_{-4854}$ \\
AT2024bgz & $43.89^{+0.01}_{-0.01}$ & $16880^{+421}_{-396}$ \\
AT2024gre & $44.13^{+0.01}_{-0.01}$ & $17125^{+356}_{-341}$ \\
\hline
\end{tabular}
\tablefoot{(1) TDE name. (2) Logarithm of the bolometric luminosity at the peak. (3) Blackbody temperature fit to the photometric data of the source at the peak.}
\end{table}

\subsection{Black hole and stellar mass estimates}
\label{massest}

To constrain the black hole mass ($M_{\rm BH}$) and the mass of the disrupted star ($M_{\rm star}$), we employed two approaches based on different publicly available codes:

\begin{itemize}
\item {\tt TDEMass} \citep{ryu2020},  which uses an outer shock model to describe the TDE emission, requiring peak $L_{\rm bol}$ and $T_{\rm bb}$ as inputs. This method assumes shocks during debris circularization are the main power source of the flare. The code is calibrated for $10^5 \la M_{\rm BH}/{\rm M_\odot} \la 5\times10^7$ and $0.1 \la M_{\rm star}/{\rm M_\odot} \la 40$; two of our events (AT2022hvp and AT2023clx) fall outside this range and are thus omitted from the corresponding columns.

\item {\tt MOSFiT} \citep{guillochon2018}, employing the fast circularization model \citep{mockler2019}, assuming that the optical and UV emission originates from X-ray photons produced by an optically thick gas layer surrounding the rapidly formed accretion disk. We fit the multi-band light curves with ten free parameters (fixing the source $z$), and derived posterior distributions for key physical properties, including $M_{\rm BH}$, $M_{\rm star}$, column density of the gas ($N_{\rm H}$) in the central region of the host galaxy, and the scaled impact parameter $b$, which is a measure of how close to the black hole the star is disrupted. 
\end{itemize}

Both approaches used host-subtracted and dereddened data as described in Section~\ref{bbfit}. The results are listed in Table~\ref{tab:masses}. Following \cite{mockler2019}, we include a systematic uncertainty of $\pm0.20$ dex to the $M_{\rm BH}$ estimate and  a systematic uncertainty of $\pm0.66$ dex on the $M_{\rm star}$ estimate. In general, \texttt{MOSFiT} returns larger $M_{\rm BH}$ values than \texttt{TDEMass}, while the stellar-mass estimates are broadly consistent. AT2022fpx and AT2022gri are especially striking: their {\tt MOSFiT}-derived $M_{\rm BH}$ values are significantly larger than those of the rest of the sample, which also translates into correspondingly large $M_{\rm star}$. In contrast, {\tt TDEMass} places both AT2022fpx and AT2022gri at $M_{\rm BH}\sim10^6$~M$\odot$ and subsolar $M_{\rm star}$. Meanwhile, AT2020mot is mostly consistent between both models within 2$\sigma$ uncertainty, and aligns reasonably well with the general population, despite having the third-largest {\tt MOSFiT}-derived $M_{\rm BH}$ value in our sample.

\begin{table*}[h!]
\renewcommand{\arraystretch}{1.25}
\caption{TDEmass and MOSFit mass results}
\label{tab:masses}
\centering
\begin{tabular}{l c c c c}
\hline\hline
Name & {\tt TDEMass} $M_{\rm BH}$ &  {\tt TDEMass} $M_*$ & {\tt MOSFiT} $M_{\rm BH}$ & {\tt MOSFiT} $M_*$ \\
 & [$10^6$ M$_\odot$] & [M$_\odot$] & [$10^6$ M$_\odot$] & [M$_\odot$] \\
(1) & (2) & (3) & (4) & (5)\\
\hline
AT2020mot & $2.80^{+0.96}_{-0.78}$ & $0.97^{+0.08}_{-0.07}$ & $18.20^{+15.68}_{-7.88}$ & $0.63^{+2.40}_{-0.56}$ \\
AT2020afhd & $0.75^{+0.17}_{-0.10}$ & $0.80^{+0.02}_{-0.02}$ & $2.95^{+3.22}_{-1.57}$ & $1.21^{+4.65}_{-0.48}$ \\ 
AT2022dbl & $1.00^{+0.24}_{-0.20}$ & $0.67^{+0.03}_{-0.03}$ & $1.20^{+0.94}_{-0.54}$ & $1.00^{+3.62}_{-0.81}$ \\ 
AT2022dbl$_2$ & $0.36^{+0.12}_{-0.09}$ & $0.23^{+0.05}_{-0.05}$ & $5.13^{+5.34}_{-2.00}$ & $0.67^{+2.55}_{-0.66}$ \\ 
AT2022fpx & $10.00^{+0.93}_{-0.89}$  & $0.92^{+0.03}_{-0.03}$ & $213.80^{+132.94}_{-81.97}$ & $14.25^{+51.55}_{-12.42}$ \\ 
AT2022gri & $1.50^{+0.31}_{-0.27}$ & $0.43^{+0.04}_{-0.04}$ & $281.84^{+196.79}_{-112.02}$ & $15.07^{+54.95}_{-13.90}$ \\ 
AT2022hvp & ...  & ... & $14.13^{+14.05}_{-6.89}$ & $0.99^{+3.59}_{-0.82}$ \\
AT2022upj & $1.30^{+0.45}_{-0.35}$ & $0.62^{+0.05}_{-0.04}$ & $4.57^{+5.90}_{-2.53}$ & $0.05^{+0.19}_{-0.04}$\\ 
AT2022wtn & $3.90^{+0.48}_{-0.45}$ & $0.50^{+0.02}_{-0.03}$ & $6.03^{+5.72}_{-3.08}$ & $0.14^{+0.59}_{-0.13}$ \\ 
AT2023clx & ... & ... & $0.47^{+5.29}_{-0.41}$ & $0.05^{+0.30}_{-0.04}$\\ 
AT2023ugy & $1.90^{+0.28}_{-0.13}$ & $2.2^{+0.45}_{-0.31}$ & $9.33^{+12.55}_{-5.61}$ & $0.52^{+3.18}_{-0.51}$\\
AT2023lli & $0.58^{+0.40}_{-0.24}$ & $0.51^{+0.07}_{-0.07}$ & $3.47^{+3.29}_{-1.69}$ & $0.40^{+1.64}_{-0.39}$\\ 
AT2024bgz & $10.00^{+0.93}_{-0.89}$ & $1.00^{+0.03}_{-0.03}$ & $2.63^{+3.26}_{-1.48}$ & $0.22^{+1.08}_{-0.21}$ \\
AT2024gre & $14.00^{+0.51}_{-0.54}$ & $2.10^{+0.22}_{-0.20}$ & $2.51^{+2.86}_{-1.22}$ & $0.57^{+2.31}_{-0.56}$ \\
\hline
\end{tabular}
\tablefoot{(1) TNS name of the source. (2) Black hole mass estimated using the {\tt TDEMass} code. (3) Mass of the disrupted star inferred using the {\tt TDEMass} code. (4) Black hole mass estimated using the {\tt MOSFiT} code. (5) Mass of the disrupted star inferred using the {\tt MOSFiT} code.}
\end{table*}

\subsection{Light curve fitting}

To estimate the rise time ($t_{\rm rise}$) of each flare, we fit the ZTF $g$ and $r$ light curves with an asymmetric Gaussian profile using a minimum-$\chi^2$ approach. The apparent magnitude $m(t)$ is modelled as:

\[
m(t) = \begin{cases}
m_{\rm host} - A \exp\left(-\frac{t-t_{\rm peak}}{2\sigma}\right)^2, & t \leq t_{\text{peak}}, \\
m_{\rm host} - B -A\exp\left(-\frac{t-t_{\rm peak}}{2\tau}\right)^2, & t > t_{\text{peak}},
\end{cases}
\]
where $m_{\rm host}$ is the median pre-flare magnitude, $A$ is the flare amplitude, and $\sigma$ and $\tau$ characterize the widths of the rise and decay phases, respectively. The parameter $B$ accounts for any plateau-like emission above the host after peak brightness \citep[e.g.,][]{mummery2024}.

We define $t_{\rm rise}$ as the difference between the times at which the flare amplitude first reaches 1\% and then 99\% of its peak, based on $\sigma$. Although this approach could underestimate the total rise if observations are sparse, it treats all sources uniformly and thus remains suitable for our population study. The $t_{\rm peak}$ and $t_{\rm rise}$ parameters determined from the light curve fitting are displayed in Table \ref{tab:trise}. Given the sparse sampling of the ZTF data for AT2023ugy and AT2023lli, the light curve fitting for those two sources has been performed using Asteroid Terrestrial-impact Last Alert System (ATLAS) c- and o- band data.

\begin{table}[h!]
\renewcommand{\arraystretch}{1.25}
\caption{Light curve fitting results}
\label{tab:trise}
\centering
\begin{tabular}{c c c}
\hline\hline
Name & $t_{\rm peak}$ & $t_{\rm rise}$ \\
 & [d] & [d] \\
(1) & (2) & (3)\\
\hline
AT2020mot & 59057.87 & $71.5\pm2.5$ \\
AT2020afhd & 60351.53 & $68.6\pm1.8$ \\
AT2022dbl & 59636.28 & $24.0\pm1.2$ \\
AT2022dbl$_2$ & 60342.72 & $52.3\pm3.4$ \\
AT2022fpx & 59785.53 & $136.4\pm1.3$ \\
AT2022gri & 59759.52 & $129.4\pm8.4$ \\
AT2022hvp & 59692.11 & $9.2\pm0.8$ \\
AT2022upj & 59887.48 & $104.0\pm4.2$ \\
AT2022wtn & 59874.31 & $35.3\pm1.4$ \\
AT2023clx & 59987.50 & $2.9\pm10.1$ \\
AT2023ugy & 60233.40 & $42.2\pm3.5$ \\
AT2023lli & 60169.69 & $67.7\pm1.9$ \\
AT2024bgz & 60349.57 & $29.0\pm2.1$ \\
AT2024gre & 60441.22 & $56.4\pm2.2$ \\
\hline
\end{tabular}
\tablefoot{(1) TDE name. (2) Peak time MJD. (3) Duration of the rise phase of the TDE flare.}
\end{table}

\subsection{X-ray luminosity}

We derived the peak X-ray luminosity $L_{\rm X}$ in the 0.3-10 keV range for each event from its X-ray flux $F_{\rm X}$, and only one source (AT2020afhd) showed detectable X-ray emission near maximum light; for the rest we report 3$\sigma$ upper limits based on non-detections.

Table~\ref{tab:compMOT} compares the key physical parameters of AT2020mot with those of the rest of the sample. For each parameter, we list the median value of the sample (excluding AT2020mot) along with its standard deviation, then contrast these with the corresponding values of AT2020mot. We find that AT2020mot is consistent with the rest of the sample for most of the measured parameters. However, it shows a marked deviation in both its maximum polarisation degree ($\Pi_{\rm max}$) and the $N_{\rm H}$ inferred from {\tt MOSFiT}.

\begin{table*}[h!]
\renewcommand{\arraystretch}{1.25}
\caption{Comparison of global properties}
\label{tab:compMOT}
\centering
\begin{tabular}{c c c c c}
\hline\hline
Parameter & Median & $\sigma$ & AT2020mot & Agreement\\
(1) & (2) & (3) & (4) & (5) \\
\hline
$T$ [K] & 29600 & 12100 & 31700 & $\checkmark$ \\ 
$\log L_{\rm bol}$ [erg s$^{-1}$] & 43.77 & 0.61 & 44.10 & $\checkmark$\\ 
$\log L_{\rm X}$ [erg s$^{-1}$] & 42.45 & 0.69 & <42.69 & $\checkmark$\\ 
$t_{\rm rise}$ [d] & 56.4 & 40.9 & 71.5 & $\checkmark$\\ 
$z$ & 0.049 & 0.036 & 0.07 & $\checkmark$ \\ 
$\Pi_{\rm max}$ [\%] & 4.1 & 2.3 & 25 & X \\ 
$\log M_{\rm BH,TDEMass}$ [M$_\odot$] & 6.18 & 0.52 & 6.45 & $\checkmark$ \\ 
$M_{\rm star,TDEMass}$ [M$_\odot$] & 0.67 & 0.62 & 0.97 & $\checkmark$ \\ 
$\log M_{\rm BH, MOSFiT}$ [M$_\odot$] & 6.66 & 0.76 & 7.26 & $\checkmark$\\ 
$M_{\rm star, MOSFiT}$ [M$_\odot$] & 0.57 & 5.11 & 0.63 & $\checkmark$\\ 
$b$ & 0.96 & 0.39 & 0.91 & $\checkmark$\\ 
$\log N_{\rm H}$ [cm$^{-2}$] & 18.29 & 1.14 & 20.80 & X\\ 
\hline
\end{tabular}
\tablefoot{(1) Parameter. (2) Median value across the sample (excluding AT2020mot). (3) Standard deviation of that parameter (excluding AT2020mot). (4) Value of AT2020mot. (5) Whether AT2020mot matches the population within 1$\sigma$ interval.}
\end{table*}

\section{Discussion}
\label{discussions}

\subsection{Polarisation trends}

We have presented new polarimetric observations of a sample of TDEs, most of which exhibit $\Pi<5\%$ or remain below our detection threshold. Such low values are consistent with expectations from both rapid–disc–formation models, in which X–rays are reprocessed by an optically thick layer \citep[e.g.][]{leloudas2022,charalampopoulos2023}, and shock-powered scenarios \citep[e.g.][]{piran2015,shiokawa2015,charalampopoulos2023}, although the latter allow for a wider range of $\Pi$ and a time–variable polarisation angle that depends on the geometry of the colliding debris streams. At present these measurements cannot discriminate between the two pictures. 
Additionally, low polarisation detection is consistent with the trends observed in other works, where polarimetric observations of TDEs reported low observed polarisation in the absence of a jet for ASASSN-18pg \citep{holoien2020}, AT2019qiz \citep{patra2022}, AT2018dyb, AT2019azh and AT2019dsg from \cite{leloudas2022}, AT2022fpx \citep{koljonen2024}, and AT2023clx \citep{koljonen2025}.
AT2020mot stands out in this context, displaying remarkably high polarisation even before any subtraction of the host flux \citep{liodakis2023}. After accounting for host contamination, $\Pi$ remains exceptionally large, despite the absence of any jet signature \citep{liodakis2023} and the lack of a concurrent X-ray detection at peak luminosity. This behaviour contrasts with the majority of our TDE sample, where polarisation remains low throughout the flare.

Once host flux is removed, AT2022hvp and AT2022upj each show a single measurement with $\Pi>6\%$, exceeding the predictions of electron scattering models \citep{leloudas2022}. Although these detections are less significant than those of AT2020mot, they demonstrate that host corrections can unveil higher polarisation states in other events. 

The two polarimetric measurements of AT2022upj obtained on the same night in the B- and R-bands hint at a strong wavelength dependence of $\Pi$. However, this is likely the result of continuum depolarisation due to the prominent H$\alpha$ line, which lies in the R band (see Appendix \ref{sampdesc}).

Several TDEs in our sample also show modest variability in $\Pi$ or $\Theta$. For instance, AT2022hvp and AT2022fpx \citep{koljonen2024}, along with AT2023clx \citep{koljonen2025}, exhibit changes in $\Pi$ during the flare. Another intriguing case is AT2024gre, which displays (within $3\sigma$) nearly constant values of both $\Pi$ and $\Theta$, albeit only over 9 days of observation. Given its featureless spectrum, one might speculate that a stable polarisation signature could be linked to geometric or optical-depth effects in the reprocessing region.

As previously mentioned, it is possible that high-$\Pi$ states may simply have been missed in most TDEs. Flares often last a year or longer; AT2020mot, for example, persisted for $\sim300$ days, yet \citet{liodakis2023} obtained only four polarimetric epochs. Assuming that each epoch samples a single night, the coverage is merely $\sim$1\% of the flare, and the extreme polarisation appears in just one night. Since other TDEs have similar fractional coverage, it is plausible that similarly strong, brief polarisation peaks have gone undetected. The physical mechanism capable of producing such extreme $\Pi$, however, remains unclear.

\subsection{Shock versus reprocessing}

When we compare AT2020mot with other TDEs (Table \ref{tab:compMOT}), most of its global parameters -e.g., $T_{\rm BB}$, $L_{\rm bol}$, $t_{\rm rise}$- fall within the typical range of these parameters for the observed TDEs. However, AT2020mot departs substantially from the norm in two key aspects: it exhibits a much higher polarisation degree than any other event in our sample, and it also appears to have an unusually large $N_{\rm H}$ in our {\tt MOSFiT} fits.

Intriguingly, \citet{Newsome2024mot} report a strong infrared flare associated with this event, which may point to a thick layer of dust or gas in the vicinity of the disruption. Normally, one might expect a substantial reprocessing layer to suppress polarisation, as scattering in optically thick material tends to reduce net polarisation signals \citep{charalampopoulos2023}.
The high polarisation of AT2020mot thus appears incompatible with this scenario unless other factors -such as tidal shocks or geometric asymmetries- are at play. Another possibility is that certain modelling assumptions (for instance, those in {\tt MOSFiT}) may not hold in this particular system, leading to an overestimated $N_{\rm H}$ or an erroneous description of the reprocessing environment. 

It is important to note that high $N_{\rm H}$ and high polarisation can indeed coexist: in Seyfert 2 nuclei high polarisation ($\Pi\sim$ 20-30\%) has been observed in the X-rays, produced from the back-scattering off material above the torus despite the high $N_{\rm H}$ that characterises these sources \citep{ursini2023,marin2024}. In the studied Seyfert 2 galaxies, however, no variability in the polarisation angle was observed, contrary to AT2020mot. The physical conditions in the TDE environment are indeed expected to be very different from Seyfert 2 galaxies, and reproducing similar polarisation behaviour would require a clumpy medium, under specific physical conditions, orbiting the SMBH. We caution that the elevated $N_{\rm H}$ inferred for AT2020mot is model dependent and could arise from features of the light curve used during the {\tt MOSFiT} fit rather than the intrinsic properties of the event. Independent diagnostics, such as the Balmer decrement, could provide an additional estimate of $N_{\rm H}$. Although the AT2020mot flare has now concluded and the available optical spectra are not suitable for similar measurements due to the low S/N, this test could possibly be applied to future TDEs via optical spectroscopic observations.

Taken together, the extreme optical polarisation, elevated $N_{\rm H}$, and strong infrared emission render AT2020mot an outlier among TDEs. Although various models can accommodate parts of the observed behaviour, no single mechanism fully explains the data. 
Polarisation variability, however, significantly disfavours reprocessing models in general, and the observed polarisation behaviour favours shock models \citep{koljonen2025}. Future theoretical investigations and more extensive polarimetric and multiwavelength monitoring of sources that exhibit similar properties will be essential to clarify whether AT2020mot represents an extreme end of typical TDE physics or if it is governed by additional processes that are not captured by standard reprocessing or shock-powered models. 

\section{Conclusions}
\label{conclusions}

We have analyzed 14 flares from 13 TDEs, presenting new polarimetric measurements obtained with multiple ground-based telescopes. We draw several key conclusions.

\begin{itemize}
    \item AT2020mot exhibits an extraordinarily high polarisation degree ($\Pi \sim 25\%$), significantly exceeding that of all other TDEs in our sample, which mostly show $\Pi < 6\%$ or lie below detection thresholds, a behaviour that is consistent with observations from other works.
   
   \item The polarization degree of AT2020mot evolves from low to high and again low values, suggesting that existing observations of other TDEs might have missed epochs of potentially higher $\Pi$.
    
    \item Several TDEs display varying $\Pi$ or $\Theta$ (e.g., AT2022hvp, AT2022fpx, AT2023clx), hinting at evolving geometries or optical depths.

    \item Current polarimetric data cannot firmly distinguish between shock-powered and reprocessing-dominated models, although high polarisation and variability disfavours reprocessing models. Additional data, especially time-resolved polarimetry near peak brightness, will be critical.

    \item Aside from its extraordinary polarisation and inferred column density  (even though the latter is a model dependent result), AT2020mot appears typical among the optical TDE population.
\end{itemize}

These findings motivate polarimetric monitoring that begins soon after discovery and continues throughout the flare’s evolution. Upcoming surveys such as LSST \citep{ivezic2019} will enlarge the TDE sample by orders of magnitude, enabling systematic polarimetric campaigns and providing the statistical leverage needed to map the diversity of TDE geometries. We are confident that larger, homogeneous datasets, combined with improved theoretical models, will ultimately clarify the physical conditions that govern TDE emission and its polarisation signature.

\begin{acknowledgements}

AF, IL, AP and BAG were funded by the European Union ERC-2022-STG - BOOTES - 101076343. Views and opinions expressed are however those of the author(s) only and do not necessarily reflect those of the European Union or the European Research Council Executive Agency. Neither the European Union nor the granting authority can be held responsible for them. KIIK has received funding from the European Research Council (ERC) under the European Union’s Horizon 2020 research and innovation programme (grant agreement No. 101002352, PI: M. Linares). The IAA-CSIC co-authors acknowledge financial support from the Spanish "Ministerio de Ciencia e Innovaci\'{o}n" (MCIN/AEI/ 10.13039/501100011033) through the Center of Excellence Severo Ochoa award for the Instituto de Astrof\'{i}sica de Andaluc\'{i}a-CSIC (CEX2021-001131-S), and through grants PID2019-107847RB-C44 and PID2022-139117NB-C44. PC acknowledges support via Research Council of Finland (grant 340613). MADT acknowledges support from the EDUFI Fellowship and the Johannes Andersen Student Programme at the Nordic Optical Telescope. The data in this study include observations made with the Nordic Optical Telescope, owned in collaboration by the University of Turku and Aarhus University, and operated jointly by Aarhus University, the University of Turku and the University of Oslo, representing Denmark, Finland and Norway, the University of Iceland and Stockholm University at the Observatorio del Roque de los Muchachos, La Palma, Spain, of the Instituto de Astrofisica de Canarias. The data presented here were obtained in part with ALFOSC, which is provided by the Instituto de Astrof\'{\i}sica de Andaluc\'{\i}a (IAA) under a joint agreement with the University of Copenhagen and NOT. Some of the data are based on observations collected at the Observatorio de Sierra Nevada; which is owned and operated by the Instituto de Astrof\'isica de Andaluc\'ia (IAA-CSIC); and at the Centro Astron\'{o}mico Hispano en Andaluc\'ia (CAHA); which is operated jointly by Junta de Andaluc\'{i}a and Consejo Superior de Investigaciones Cient\'{i}ficas (IAA-CSIC). We acknowledge funding to support our NOT observations from the Finnish Centre for Astronomy with ESO (FINCA), University of Turku, Finland (Academy of Finland grant nr 306531). E.L. was supported by Academy of Finland projects 317636 and 320045. This research has made use of data from the RoboPol programme, a collaboration between Caltech, the University of Crete, IA-FORTH, IUCAA, the MPIfR, and the Nicolaus Copernicus University, which was conducted at Skinakas Observatory in Crete, Greece. 

\end{acknowledgements}

%
%

\bibliographystyle{aa}
\bibliography{biblio.bib}

\begin{appendix}

\onecolumn
\section{New observations}

In this section we display the results from the polarimetric observations of the TDEs employed in this work as displayed in Table \ref{tab:newpol}. In Table \ref{tab:hostcorr} we display the host-corrected polarisation observations, performed following the criteria described in Section \ref{hostcorr}.

\begin{table*}[h!]
\renewcommand{\arraystretch}{1.10}
\caption{New polarisation observations}
\label{tab:newpol}
\centering
\begin{tabular}{c c c c c c c c c}
\hline\hline
Name & MJD & $\Delta t_{\rm peak}$ & $\Pi$ & $\Theta$ & q & u & Telescope & Filter  \\
 & [d] & [d] & [\%] & [$^\circ$] & [\%] & [\%] & & \\
(1) & (2) & (3) & (4) & (5) & (6) & (7) & (8) & (9)\\
\hline
AT2020afhd & 60345.351 & -6.18 & $<1.8$ & ... & $1.44\pm0.60$ & $-0.40\pm0.60$ & OSN T90 & R \\
 & 60345.384 & -6.15 & $<1.8$ & ... & $-0.73\pm0.61$ & $1.08\pm0.61$ & OSN T90 & R\\
 & 60347.384 & -4.15 & $<2.1$ & ... & $1.56\pm0.70$ & $0.66\pm0.70$ & OSN T90 & R\\
 & 60348.324 & -3.21 & $<1.8$ & ... & $-0.10\pm0.61$ & $-0.03\pm0.72$ & OSN T90 & R\\ 
 & 60371.319 & +19.79 & $<2.1$ & ... & $0.16\pm0.70$ & $0.78\pm0.70$ & OSN T90 & R\\
 & 60384.333 & +32.80 & $<2.7$ & ... & $-0.05\pm0.99$ & $-0.40\pm0.90$ & OSN T90 & R\\
\hline
AT2020mot & 59867.567 & +809.70 & $<0.75$ & ... & $0.53\pm0.24$ & $0.59\pm0.25$ & NOT & R\\ 
 & 59896.531 & +838.66 & $<0.7$ & ... & $-0.19\pm0.29$ & $-0.35\pm0.30$ & NOT & R\\ 
 & 59929.465 & +871.60 & $<0.45$ & ... & $-0.17\pm0.15$ & $0.22\pm0.15$ & NOT & R\\ 
 & 59936.472 & +878.60 & $<0.42$ & ... & $0.04\pm0.14$ & $0.24\pm0.14$ & NOT & R\\ 
 & 60180.649 & +1122.78 & $<0.44$ & ... & $-0.14\pm0.27$ & $-0.11\pm0.26$ & NOT & R\\ 
 & 60207.492 & +1149.62 & $<0.33$ & ... & $-0.05\pm0.11$ & $0.34\pm0.11$ & NOT & V\\ 
 & 60207.583 & +1149.71 & $<0.35$ & ... & $-0.20\pm0.15$ & $-0.03\pm0.15$ & NOT & R\\ 
\hline
AT2022dbl & 59647.578 & +11.30 & $<0.51$ & ... & $-0.01\pm0.20$ & $0.29\pm0.16$ & NOT & R\\ 
\hline
AT2022dbl$_2$ & 60357.720 & +12.93 & $<2.1$ & ... & $-0.09\pm0.66$ & $-0.01\pm0.63$ & OSN T90 & R\\
 & 60385.660 & +40.87 & $<1.8$ & ... & $-0.22\pm0.59$ & $0.03\pm0.60$ & OSN T90 & R\\ 
 & 60402.488 & +57.70 & $<1.8$ & ... & $-0.01\pm0.65$ & $-0.24\pm0.63$ & OSN T90 & R\\ 
 & 60411.462 & +66.67 & $<2.1$ & ... & $0.03\pm0.71$ & $0.74\pm0.70$ & OSN T90 & R\\ 
 & 60426.308 & +81.52 & $<1.8$ & ... & $1.10\pm0.67$ & $-0.70\pm0.37$ & Skinakas 1.3m & r\\ 
 \hline
AT2022gri & 59720.403 & -39.12 & $0.65\pm0.15$ & $162.6\pm6.8$ & $0.55\pm0.16$ & $-0.38\pm0.17$ & NOT & B \\ 
\hline
AT2022hvp & 59713.327 & +21.22 & $4.39\pm0.71$ & $47.5\pm4.2$ & $-0.39\pm0.64$ & $4.37\pm0.71$ & Skinakas 1.3m & R\\ 
 & 59738.336 & +46.23 & $<2.79$ & ... & $-0.22\pm0.94$ & $0.30\pm0.93$ & Skinakas 1.3m & R\\ 
\hline
AT2022upj & 59909.480 & +22.00 & $<0.45$ & ... & $-0.26\pm0.17$ & $-0.07\pm0.17$ & NOT & R\\ 
 & 59909.496 & +22.02 & $2.92\pm0.26$ & $48.4\pm2.5$ & $-0.37\pm0.27$ & $3.11\pm0.24$ & NOT & B\\ 
\hline
AT2022wtn & 59930.427 & +56.12 & $<0.72$ & ... & $-0.58\pm0.28$ & $0.40\pm0.28$ & NOT & B\\ 
 & 59930.442 & +56.13 & $<0.48$ & ... & $-0.40\pm0.18$ & $-0.22\pm0.18$ & NOT & R\\ 
\hline
AT2023ugy & 60240.324 & +6.92 & $<1.9$ & ... & $0.82\pm0.61$ & $-0.60\pm0.71$ & Skinakas 1.3m & r\\
& 60309.348 & +75.95 & $<1.2$ & ... & $0.82\pm0.48$ & $0.85\pm0.48$ & NOT & B\\ 
& 60309.362 & +75.96 & $<0.75$ & ... & $0.24\pm0.29$ & $0.51\pm0.29$ & NOT & R\\ 
\hline
AT2023lli & 60207.620 & +37.93 & $0.62\pm0.20$ & $115.1\pm10.9$ & $-0.42\pm0.23$ & $-0.51\pm0.23$ & NOT & B\\ 
 & 60207.635 & +37.95 & $<0.48$ & ... & $-0.03\pm0.16$ & $0.02\pm0.16$ & NOT & R\\ 
 & 60208.490 & +38.80 & $<0.42$ & ... & $0.26\pm0.14$ & $0.07\pm0.13$ & Skinakas 1.3m & r\\ 
 & 60237.344 & +67.65 & $0.50\pm0.13$ & $45.5\pm10.3$ & $-0.01\pm0.18$ & $0.50\pm0.13$ & Skinakas 1.3m & r\\ 
 & 60250.469 & +80.78 & $<0.57$ & ... & $-0.19\pm0.22$ & $-0.44\pm0.22$ & NOT & R\\ 
\hline
AT2024bgz & 60371.411 & +21.84 & $<5.4$ & ... & $1.00\pm1.85$ & $-2.29\pm1.81$ & OSN T90 & R\\
 & 60385.561 & +35.99 & $<8.1$ & ... & $2.57\pm2.69$ & $-0.44\pm2.47$ & OSN T90 & R\\ 
 & 60404.416 & +54.85 & $<7.8$ & ... & $4.38\pm2.60$ & $-0.44\pm2.47$ & OSN T90 & R\\ 
\hline
AT2024gre & 60470.416 & +29.20 & $2.43\pm0.37$ & $75.3\pm4.4$ & $-2.12\pm0.38$ & $1.20\pm0.36$ & NOT & B\\ 
 & 60470.431 & +29.21 & $1.32\pm0.29$ & $83.1\pm6.5$ & $-1.31\pm0.32$ & $0.32\pm0.32$ & NOT & R\\ 
 & 60478.404 & +37.18 & $2.05\pm0.63$ & $64.7\pm9.9$ & $-1.38\pm0.71$ & $1.68\pm0.74$ & NOT & B\\ 
 & 60478.419 & +37.20 & $1.48\pm0.42$ & $84.1\pm9.0$ & $-1.53\pm0.48$ & $0.32\pm0.49$ & NOT & R\\ 
 & 60479.381 & +38.16 & $2.5\pm0.6$ & $79.7\pm6.2$ & $-2.34\pm0.59$ & $0.88\pm0.55$ & CAHA 2.2m & R\\
 & 60479.426 & +38.21 & $2.0\pm0.6$ & $65.3\pm8.4$ & $-1.30\pm0.59$ & $1.52\pm0.59$ & CAHA 2.2m & R\\ 
\hline
\end{tabular}
\tablefoot{(1) TDE name. (2) MJD of the observation. (3) Time from peak. (4) Measured polarisation degree, with the associated uncertainty. (5) Measured polarisation angle, with the associated uncertainty. The measured polarisation angle is only reported when the polarisation degree is detected within 3$\sigma$. (6) Measured q Stokes parameter, with the associated uncertainty. (7) Measured u Stokes parameter, with the associated uncertainty. (8) Telescope with which the observation was conducted. (9) Observing filter.}
\end{table*}

\begin{table*}[h!]
\renewcommand{\arraystretch}{1.10}
\caption{Host-corrected observations}
\label{tab:hostcorr}
\centering
\begin{tabular}{c c c c c c c}
\hline\hline
Name & MJD & $\Delta t_{\rm peak}$ & $\Pi_{\rm corr}$ & $\Theta$ & Telescope & Filter  \\
 & [d] & [d] & [\%] & [$^\circ$]& & \\
(1) & (2) & (3) & (4) & (5) & (6) & (7)\\
\hline
AT2022gri & 59720.403 & -39.12 & $1.29\pm0.39$ & $162.6\pm6.8$ & NOT & B \\ 
\hline
AT2022hvp & 59713.327 & +21.22 & $6.86\pm1.14$ & $47.5\pm4.2$ & Skinakas 1.3m & R\\ 
 & 59738.336 & +46.23 & $<6.34$ & ... & Skinakas 1.3m & R\\ 
\hline
AT2022upj & 59909.480 & +22.00 & $<2.34$ & ... & NOT & R\\ 
 & 59909.496 & +22.02 & $6.40\pm1.20$ & $48.4\pm2.5$ & NOT & B\\ 
\hline
AT2024gre & 60470.416 & +29.20 & $2.74\pm0.43$ & $75.3\pm4.4$ & NOT & B\\ 
 & 60470.431 & +29.21 & $1.97\pm0.48$ & $83.1\pm6.5$ & NOT & R\\ 
 & 60478.404 & +37.18 & $2.32\pm0.72$ & $64.7\pm9.9$ & NOT & B\\ 
 & 60478.419 & +37.20 & $2.39\pm0.75$ & $84.1\pm9.0$ & NOT & R\\ 
 & 60479.381 & +38.16 & $4.1\pm1.1$ & $79.7\pm6.2$ &  CAHA 2.2m & R\\
 & 60479.426 & +38.21 & $3.2\pm1.1$ & $65.3\pm8.4$ &  CAHA 2.2m & R\\ 
\hline
\end{tabular}
\tablefoot{(1) TDE name. (2) MJD of the observation. (3) Time from peak. (4) Host-corrected polarisation degree, with the associated uncertainty. (5) Measured polarisation angle, with the associated uncertainty. The measured polarisation angle is only reported when the polarisation degree is detected within 3$\sigma$. (6) Telescope with which the observation was conducted. (7) Observing filter.}
\end{table*}

\twocolumn

\section{Sample description}
\label{sampdesc}

In this section, we briefly describe each source in our sample, summarizing relevant findings from previous studies and providing details on their spectral classification. We also discuss their X-ray behaviour and host galaxy classification, where available.\\

{\bf AT2020mot} was first discovered on 14 June 2020, and identified as a TDE H+He due to the presence of the broad He{\sc{ii}}$\lambda$4686 line together with H$\beta$ in its spectrum \citep{hosseinzadeh2020}. RoboPol observations obtained with the 1.30m telescope at Skinakas observatory revealed its unique nature: it exhibits the highest measured polarisation degree ($\Pi \sim 25\%$), consistent across the $V$, $R$, and $I$ bands after subtracting the unpolarised host flux contribution \citep{liodakis2023}. Polarimetric observations started $\sim 25$ days after the optical peak and continued for nearly three years. \citet{liodakis2023} also reports radio observations with the Very Large Array (VLA) at 15 GHz, yielding a 27 $\mu$Jy upper limit non-detection. \citet{Newsome2024mot} report a strong $i$-band excess, interpretable as emission from two concentric dust rings. AT2020mot does not exhibit X-ray emission across the entire duration of the TDE flare. Its host, WISEA,J003113.52+850031.8 (at $z=0.07$), is classified as an E+A galaxy \citep{hosseinzadeh2020}, a type commonly associated with TDEs \citep{hammerstein2021}.\\

{\bf AT2020afhd} was initially discovered on 20 October 2020, but exhibited an optical flare in 2024 that classifies it as a TDE H+He due to its strong Balmer and He{\sc{ii}} emission \citep{hammerstein2024} and a Bowen fluorescence flare (BFF, \citealt{arcavi2024}).  Polarimetry covers from $\sim6$ days before to $\sim30$ days after peak. AT2020afhd exhibited X-ray emission at peak luminosity and the emission persisted during the decay phase of the TDE flare.
VLA observations at 15 GHz detected a flux density of $253\pm18$ $\mu$Jy at the transient coordinates \citep{christy2024}. Its host, LEDA 145386 (at $z=0.027$), is classified as a Seyfert 2 galaxy.\\

{\bf AT2022gri} was discovered on 3 April 2022 and identified as a featureless TDE, showing a blue continuum superimposed on the host-galaxy absorption lines at $z=0.028$ \citep{yao2022}. A single polarimetric epoch with ALFOSC@NOT was obtained $\sim40$ days before peak. No X-ray emission has been detected.\\ 

{\bf AT2022hvp} was discovered on 19 April 2022 and classified as a TDE He due to the broad He{\sc{ii}} line on top of a blue continuum \citep{fulton2022}. Polarimetry spans from $\sim20$ to $\sim45$ days after peak luminosity. AT2022hvp did not exhibit X-ray emission throughout the duration of the optical flare. Its host galaxy, SDSS J095445.24+552625.2 (at $z=0.12$), appears to be a red, quiescent system lacking prominent absorption features \citep{fulton2022}.\\

{\bf AT2022fpx} was discovered on 31 March 2022 and classified as a TDE H+He, given the presence of Balmer and He{\sc{ii}} lines in its spectra. The source also shows multiple high-ionization lines, classifying it as an extreme coronal line emitter (ECLE) \citep{koljonen2024}. Polarimetric observations started $\sim 10$ days after peak luminosity, and continued until $\sim 390$ days after peak luminosity. Its host galaxy (at $z=0.0735$) is an E+A galaxy with a $\sim56\%$ probability of hosting an AGN \citep{koljonen2024}.\\

{\bf AT2022upj} was first discovered on 18 September 2022 and identified as a TDE He based on the presence of the broad He{\sc{ii}} line \citep{newsome2022}. AT2022upj also exhibited the [Fe{\sc{xiv}}]$\lambda5303$ and [Fe{\sc{x}}]$\lambda6375$ high ionization lines, which are characteristic of the ECLE class of TDEs \citep{Newsome2024upj}. Although the source is X-ray bright 350 days after maximum, the preceding observation was 130 days earlier, so the onset time is uncertain. \cite{Newsome2024upj} reports X-ray detections at the time of the peak. However, the discrepancy likely arises from the more conservative S/N cut that we adopt for the events. Polarimetry was obtained $\sim20$ days after peak luminosity. Its host galaxy, LEDA 924801, is found at $z=0.054$. The source exhibits a clear discrepancy between the observed polarisation in the B- and R- bands, hinting at the possibility of a wavelength dependence. It is likely, however, that the difference is a result of dilution of the continuum polarisation from the H$\alpha$ line which is quite prominent at the time the classification spectrum was taken (see spectra at \url{https://www.wis-tns.org/object/2022upj}). Since no contemporary spectra were taken at the time of our polarisation measurements, a quantitative correction is not feasible. \\

{\bf AT2022wtn} was discovered on 2 October 2022 and classified as a TDE H+He \citep{fulton2022wtn, onori2025}. Polarimetric observations of the source were conducted $\sim 60$ days after peak luminosity. AT2022wtn does not exhibit X-ray emission across the entire duration of the flare.
The host galaxy of AT2022wtn, SDSS J232323.79+104107.7 (at $z=0.049$), is considered to be likely hosting star-formation, and is currently merging with the more massive neighbouring galaxy SDSS J232323.37+104101.7 \citep{onori2025}. \\

{\bf AT2022dbl} was first detected on 22 February 2022 and identified as a TDE H+He based on its nuclear location and the presence of broad Balmer lines as well as He{\sc{ii}}\citep{arcavi2022dbl}. A second flare occurred at the end of 2023 \citep{yao2024}, attributed to the partial disruption of the same star, making it a partial TDE \citep[pTDE;][]{lin2024}. Polarimetric observations were conducted $\sim 7$ days after peak brightness of the first flare, and restarted $\sim 15$ days after the second flare peaked, until $\sim 80$ days after peak brightness of the second TDE flare. No X-ray emission has been detected throughout the duration of the flare.
AT2022dbl was also observed with the VLA at 15,GHz, yielding a flux density of $32\pm7 \mu$Jy \citep{sfaradi2022}. Its host, WISEA J122045.05+493304.7 (at $z=0.0284$), is classified as an E+A galaxy \citep{arcavi2022dbl}.\\

{\bf AT2023clx} was first detected on 22 February 2023 and classified as a TDE H+He given its nuclear position and its broad Balmer and He{\sc{ii}} lines lines superimposed on a blue continuum \citep{taguchi2023}. \citet{charalampopoulos2024} carried out a detailed spectral analysis, identifying multiple components in the emission lines. Polarimetric observations were conducted $\sim 5$ days after peak luminosity, and continued until $\sim 35$ days after peak luminosity. The source did not exhibit X-ray emission throughout the duration of the optical flare.
AT2023clx was also observed with the Arcminute Microkelvin Imager - Large Array (AMI-LA) at a central frequency of 15.5 GHz, and a flux of $0.40\pm0.08$ mJy was detected, although the transient nature of the detected signal was not confirmed \citep{sfaradi2023}.
The host galaxy, NGC 3799 (at $z=0.01107$), is an SAB LINER system that is interacting with the neighbouring galaxy NGC 3800 \citep{charalampopoulos2024}.\\

{\bf AT2023ugy} was discovered on 6 October 2023 and identified as a TDE H from its broad H$\alpha$ emission atop a blue continuum \citep{yao2023}. Polarimetric observations were conducted at peak luminosity, until $\sim 70$ days after. X-ray emission was not detected from the source throughout the TDE flare.
The host galaxy of AT2023ugy, SDSS J214044.01+210058.4 is found at $z=0.106$.\\ 

{\bf AT2023lli} was found on 23 June 2023 and classified as a TDE H+He characterized by a very broad H$\alpha$ line (FWHM~$\sim~15000$ km s$^{-1}$) and the He{\sc{ii}} line\citep{hinkle2023}. Polarimetric observations were conducted $\sim 50$ days after peak luminosity, and ended $\sim 45$ days later. X-ray emission from the source was not detected at the time of the peak but was detected starting from $\sim 110$ days after optical peak luminosity. \citet{huang2024} report $u-r=2.08$ mag for the host galaxy (at $z=0.036$), placing it in the “green valley” typical of TDE hosts \citep{hammerstein2021}, consistent with observed strong Balmer absorption features indicative of a post-starburst galaxy. AT2023lli has a predicted maximum interstellar polarisation higher than the observed polarisation, making it the only source in our sample with a significant $\Pi_{\rm max}$ ($\Pi_{\rm max}\sim0.9$\%). However, the observed variation in $\Theta$ (see Table \ref{tab:newpol}) points to an intrinsic origin to the TDE.\\

{\bf AT2024bgz} was first discovered on 1 February 2024 and identified as a TDE H+He \citep{godson2024}. VLA observations at 15 GHz did not detect any emission at the transient coordinates \citep{golay2024}. Polarimetric observations started approximately at peak luminosity, and ended $\sim 70$ days after peak luminosity. X-ray emission was not detected from the source throughout the TDE flare. The host galaxy of AT 2024bgz, 2MASX J09440480-0412051, is found at $z=0.0585$.\\ 

{\bf AT2024gre} was discovered on 16 April 2024 and identified as a featureless TDE, showing a blue continuum with no prominent emission lines \citep{somalwar2024}. Polarimetric observations started $\sim 20$ days after peak luminosity, and ended 9 days later. X-ray emission was not detected from the source throughout the TDE flare. The host galaxy of AT 2024gre, SDSS J103138.88+345430.0, is found at $z=0.12$.\\ 

\end{appendix}

\end{document}